\documentclass[useAMS,usenatbib]{mn2e}

\usepackage{latexsym}
\usepackage{amssymb}
\usepackage[dvips]{graphicx}
\usepackage{mathrsfs}

\newcommand{\mbh}{$M_{\rmn{BH}}$}

\newcommand{\mmbh}{M_{\rmn{BH}}}

\newcommand{\mmsun}{\rmn{M}_{\odot}}

\newcommand{\CIV}{C\textsc{iv}}
\newcommand{\hbeta}{H$\beta$}
\newcommand{\etal}{et al.}
\newcommand{\MgII}{Mg\textsc{ii}}
\newcommand{\Mg}{Mg}

\title[Downsizing of supermassive black holes from the SDSS quasar survey]{Downsizing of supermassive black holes from the SDSS quasar survey}
\author[M. Labita, R. Decarli, A. Treves and R. Falomo]{M. Labita$^{1}$\thanks{E-mail:
marzia.labita@gmail.com}, R. Decarli$^{1}$, A. Treves$^{1}$ and R. Falomo$^{2}$\\
$^{1}$Department of Physics and Mathematics, University of Insubria, Via Valleggio 11, I-22100 Como, Italy\\
$^{2}$INAF, Astronomical Observatory of Padova, Vicolo dell'Osservatorio 5, I-35122 Padova, Italy}
\begin{document}
\date{Accepted ... Received ...; in original form ...}
\pagerange{\pageref{firstpage}--\pageref{lastpage}} \pubyear{0000}
\maketitle
\label{firstpage}
\begin{abstract}
Starting from the $\sim$50000 quasars of the Sloan Digital Sky Survey for which \MgII\ line width and 3000 \AA\ monochromatic flux are available, we aim to study the dependence of the mass of active black holes on redshift. 
We focus on the observed distribution in the FWHM--nuclear luminosity plane, which can be reproduced at all redshifts assuming a limiting \mbh, a maximum Eddington ratio and a minimum luminosity (due to the survey flux limit). We study the $z$-dependence of the best fit parameters of assumed distributions at increasing redshift and find that the maximum mass of the quasar population evolves as $\log (M_{\rm BH\,(max)}/{\rm M}_{\odot})\sim 0.3z+9$, while the maximum Eddington ratio ($\sim0.45$) is practically independent of cosmic time.  These results are 
unaffected by the Malmquist bias.
\end{abstract}
\section{Introduction}
In the last years a substantial effort has been devoted to measure black hole (BH) masses for various quasar samples covering a wide range of redshifts and luminosities. 
McLure \& Dunlop (2004), from \hbeta\ and \MgII, measured virial black hole masses (\mbh) for $\sim10000$ quasars with $z\le 2.1$ included in the Sloan Digital Sky Survey (SDSS) Data Release 1 (DR1). 
Fine \etal\ (2006) used composite spectra to measure the dependence on redshift of the mean BH mass for an $L^*$ subsample of the 2QZ quasar catalogue (Croom \etal\ 2004) from $z\sim 0.5$ to $z\sim 2.5$.
Shen et al. (2008) listed BH masses for $\sim 60000$ quasars in the redshift range $0.1\lesssim z \lesssim4.5$ contained in the SDSS DR5, by means of virial BH mass estimators based on the \hbeta, \MgII\ and \CIV\ lines.

A common result of these works is that the mean BH mass of the QSO population at given $z$ appears to increase with redshift, but the observed $z$-dependence is dominated by the well-known Malmquist bias, because the BH mass strongly correlates with the central source luminosity (see Vestergaard \etal\ 2008 for a detailed analysis of the selection bias effects). McLure \& Dunlop (2004), for instance, suggest that the observed active BH mass evolution is entirely due to the effective flux limit of the sample. 

A full understanding of this scenario would give important insights on the BH formation and evolution and on the activation of the quasar phenomenon. Moreover, along with a parallel study on the dependence on redshift of the host galaxy luminosity (mass), this would enlighten on the joint evolution of galaxy bulges and their central black holes. For these reasons, it is of focal importance to trace the dependence of \mbh\ on $z$, overcoming the problems related to the Malmquist bias. 

We start from the recently published SDSS DR5 quasar catalog (Schneider \etal\ 2007) and focus on the $\sim50000$ quasars for which \MgII\ line width and 3000 \AA\ flux are available (Shen et al. 2008). The sample selection is described in Section \ref{sample}. The sample ($0.35<z<2.25$) is divided in 8 redshift bins and it is shown that, in each bin, the object distribution in the FWHM--luminosity plane can be reproduced assuming a minimum luminosity, a maximum mass and a maximum Eddington ratio (Sections \ref{secdistrib} and \ref{secdensity}). Comparing the assumed probability density to the observed distribution of objects, the parameters can be determined in each redshift bin (Section \ref{secfit}). 
This procedure is shown to be unaffected by the Malmquist bias (Section \ref{robust}), and provides a method to study the ``unbiassed'' dependence on redshift of quasar BH masses and Eddington ratios (Section \ref{secevoz}). In Section \ref{seclll} we test the dependence of our results on the assumed $r_{\rm BLR}-\lambda L_{\lambda}$ calibration. 
We compare our results with previous literature in Section \ref{seccomp} and in Section \ref{secdisc} we discuss some  implications of our findings for the study of the co-evolution of supermassive BHs and their host galaxies.
A summary of the paper is given in the last Section.

Throughout this paper, we adopt a concordant cosmology with $H_0=70$ ~km~s$^{-1}$~Mpc$^{-1}$, $\Omega_m=0.3$ and $\Omega_{\Lambda}=0.7$. 
\section{The \Mg II sample}\label{sample}
The SDSS DR5 quasar catalogue (Schneider et al. 2007) contains more than $77,000$ quasars. It covers about 8000 deg$^{2}$ and  selects objects with $M_i<-22$, have at least one emission line with  FWHM larger than 1000 km$/$s  or  are unambiguously  broad absorption line objects, are fainter than $i=15.0$
and have highly reliable redshifts. 

Shen et al. (2008) calculated BH masses for $\sim 60000$ quasars in the redshift range $0.1<z<4.5$ included in the SDSS DR5 quasar catalogue, using virial BH mass estimators based on the \hbeta, \MgII\ and \CIV\ emission lines. They provide rest-frame line widths and monochromatic luminosities at 5100 \AA, 3000 \AA\ and 1350 \AA\ (see Shen et al. 2008 for details on calibrations, measure procedures, corrections). 

In the following we will focus on the $\sim50000$ quasars from the Shen et al.~(2008) sample for which \MgII\ line width and 3000 \AA\ monochromatic flux are available (hereafter \MgII\ sample). We assume the virial theorem and adopt the calibration of McLure \& Dunlop (2004) to evaluate the BH mass:
\begin{equation}\label{formulamassa}
\log\mmbh{}=6+\log(a)+2\log({\rmn{FWHM}})+b\log{\lambda L_{\lambda}}
\end{equation} 
with $a=3.2\pm 1.1$ and $b=0.62\pm 0.14$. Here \mbh\ is expressed in solar masses, FWHM in units of 1000 km/s and $\lambda L_{\lambda}$ in units of $10^{44}$ erg/s.
\section{Description of the procedure}\label{secprocedure} 
\subsection{Malmquist bias}\label{secbias}
\begin{figure}
\includegraphics[width=0.37\textwidth, angle=270]{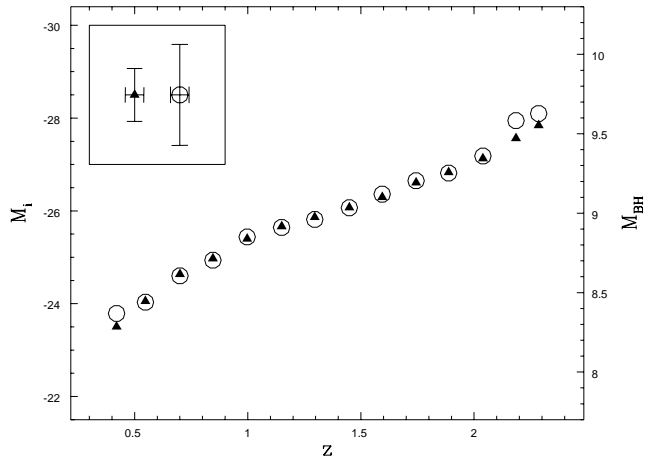}  
\caption[]{Average values of the absolute magnitude $M_i$ (triangles) and of \mbh\ (circles) vs. redshift, in bins $\Delta z=0.15$. The typical standard deviation of each bin is given in the inset.
}
\label{biasbis}
\end{figure}
Figure \ref{biasbis} shows the mean absolute $i$ magnitude (see Shen et al. 2008 for details) vs.~redshift of the \MgII\ sample. The effects of the Malmquist bias are apparent as an increase of the average observed luminosity with redshift. The mean BH mass vs.~redshift is overplotted to the mean $M_i(z)$: it is apparent that the average observed BH masses follow the same trend as the absolute magnitudes with a higher dispersion, as expected given that the distribution of the line widths does not depend on luminosity or redshift (Shen et al. 2008).
This suggests that the $z$-dependence of the observed BH masses is strongly subject to a Malmquist-type bias, because at high redshift one cannot observe low mass objects. In order to trace the ``unbiassed'' dependence of active BH masses with redshift, one  should consider a combination of two effects, namely the $z$-dependence of the quasar number density and the increase of the average mass of quasar populations with redshift.  To illustrate these effects consider two extreme cases: 
\begin{enumerate}  \item The \mbh\ distribution does not depend on redshift, but the quasar number density increases until $z\sim2-2.5$. At any redshift there is a population of low mass ($\sim10^8\mmsun$) quasars, which cannot be observed at high redshift, and the population of high mass ($\sim10^{9.5}\mmsun$) active BHs that is observed at $z\gtrsim1.5$ is the high mass end of the \mbh\ distribution. \item The quasar \mbh\ distribution shifts toward higher masses at increasing redshift. The population of low mass  ($\sim10^8\mmsun$) objects that is observed at low redshift is not present at all at $z\gtrsim1.5$. The observed increase of \mbh\ with redshift, in active BHs, is ``true'' and it is not due to a Malmquist-type bias.
\end{enumerate}
Of course, each of these pictures is {\it per se} unlikely: the observed dependence on redshift of quasar BH masses is due to a combination of both these effects. 
In the following, we will concentrate on these points using statistical arguments, focusing on the distribution of objects in the FWHM--luminosity plane.
\subsection{Quasar distribution in the FWHM--$\lambda L_{\lambda}$ plane}\label{secdistrib}
\begin{figure*}
\begin{minipage}{\textwidth}
\centering
\includegraphics[width=0.6\textwidth, angle=270]{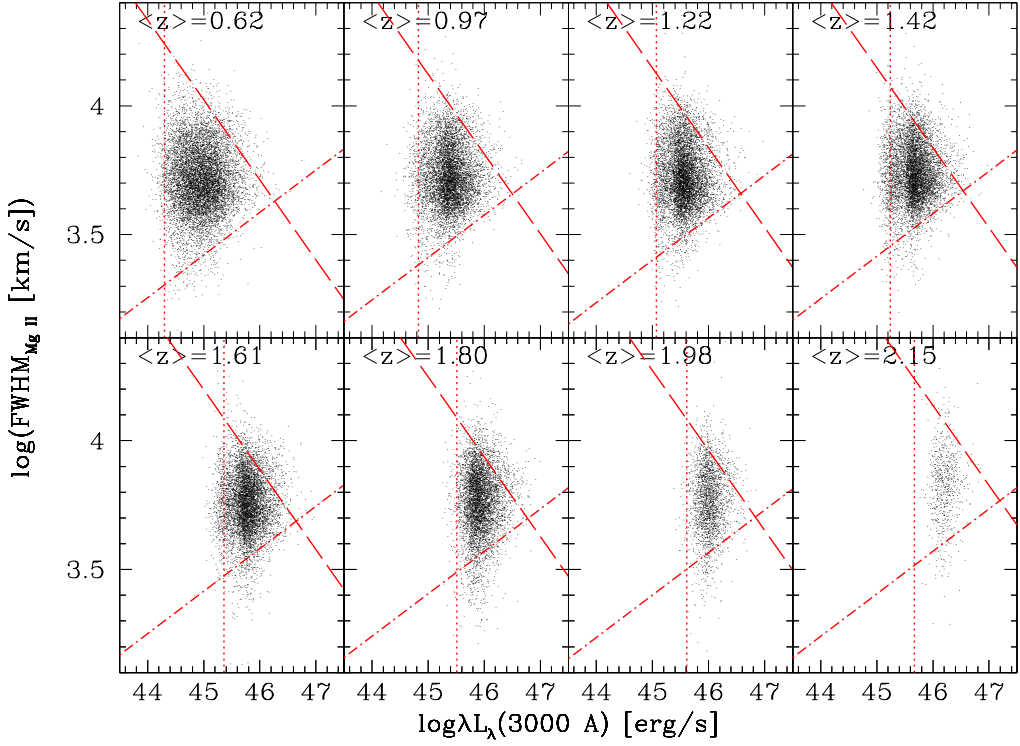}
\caption[]{The 8 panels show the \MgII\ sample in the FWHM--luminosity plane at increasing redshift. Dotted, dashed and dash-dotted lines (and lines parallel to them) represent the loci of constant monochromatic luminosity, constant mass and constant Eddington ratio respectively.}
\label{graficozsemplice}
\end{minipage}
\end{figure*}
Figure \ref{graficozsemplice} shows the objects of the \MgII\ sample in the logFWHM--log$\lambda L_{\lambda}$ plane  (see Fine et al.~2008 for a similar approach). The sample has been divided in 8 redshift bins of equal co-moving volume. In each panel, it is apparent that the data-points form a sort of ``triangle'', the left side of which represents a cut due to the the survey flux limit (which gives raise to the Malmquist bias). 
From Eq.~\ref{formulamassa}, the loci of quasars with constant mass are represented in this plane by straight lines with fixed slope, as plotted in the figure:
\begin{equation}\label{formulamassa2}
\log({\rmn{FWHM}})=-0.31\log{\lambda L_{\lambda}}+0.5\log\frac{M_{\rm BH}}{M_{\odot}}-3.25
\end{equation}
where units are the same as in Eq. \ref{formulamassa}.
We propose that the top right side of the triangle is representative of a maximum mass in the quasar sample.

The third (i.e. the bottom right) side of the triangle is supposedly due to the Eddington limit, as the loci of quasars with constant Eddington ratios are again straight lines. The dependence of FWHM on the monochromatic luminosity at a given Eddington ratio is fixed assuming the bolometric correction by Richards et al. (2006; BC$_{3000}=5.15$) and Eq.~\ref{formulamassa}. This yields:
\begin{equation}\label{formulaedd}
\log({\rmn{FWHM}})=0.19\log{\lambda L_{\lambda}}-0.5\log\frac{L_{\rm bol}}{L_{\rm Edd}}+0.05
\end{equation}
where units are the same as in Eq. \ref{formulamassa}.
Note that in each redshift bin, the plotted cuts describe qualitatively well the shape of the quasar distribution in the FWHM--luminosity plane. 
\subsection{Construction of a probability density}\label{secdensity}
\begin{figure}
\includegraphics[width=0.375\textwidth, angle=270]{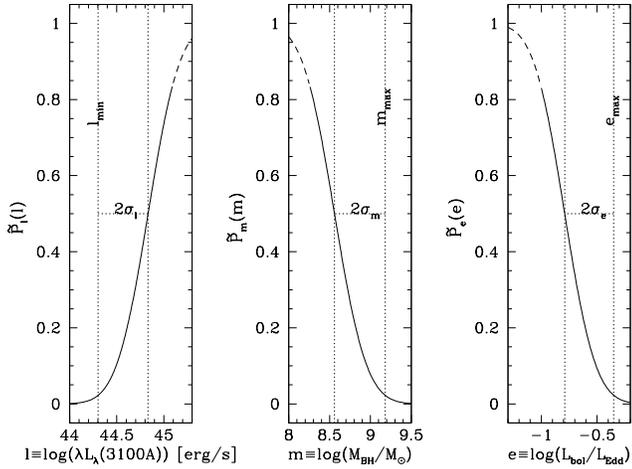}
\caption[]{The shape of $\tilde{P}_l(l)$ (Eq. \ref{PL}), $\tilde{P}_m(m)$ (Eq. \ref{PM}) and $\tilde{P}_e(e)$ (Eq. \ref{PE}).}
\label{distrib}
\end{figure}
Now we aim to construct a probability density of quasars as a function of FWHM and luminosity with   a main criterion of  simplicity. We propose that in each redshift bin the object density is only constrained by a maximum mass, a maximum Eddington ratio and a minimum luminosity due to the instrumental flux limit. 
We then assume a probability density of form:
\begin{equation}\label{Ptot}
P_{\,l,\,{\rm FWHM}}(l,\,{\rm FWHM})=k\cdot\tilde{P}_l(l)\cdot \tilde{P}_m(m)\cdot \tilde{P}_e(e)
\end{equation}
where $k$ is a normalization constant and each $\tilde{P}$ is assumed to be a smoothed step function,  which increases from 0 to 1 (or {\it vice versa}) in a range of width $\sigma$ around a fixed value of the independent variable. 
In the following, we describe our results assuming $\tilde{P}$ of form (see Figure \ref{distrib}):
\begin{equation}\label{PL}
\tilde{P}_l(l)=\frac{1}{\sigma_l \sqrt{2\pi}}\int_{-\infty}^{l}\!\!\!\!{\rm exp}\Big[-\frac{(l'-l_{\rm min}-2\sigma_l)^2}{2\sigma_l^2}\Big]dl'
\end{equation}
\begin{equation}\label{PM}
\tilde{P}_m(m)=\frac{1}{\sigma_m \sqrt{2\pi}}\int_{m}^{+\infty}\!\!\!\!\!\!{\rm exp}\Big[-\frac{(m'-m_{\rm max}+2\sigma_m)^2}{2\sigma_m^2}\Big]dm'
\end{equation}
\begin{equation}\label{PE}
\tilde{P}_e(e)=\frac{1}{\sigma_e \sqrt{2\pi}}\int_{e}^{+\infty}\!\!\!\!\!\!{\rm exp}\Big[-\frac{(e'-e_{\rm max}+2\sigma_e)^2}{2\sigma_e^2}\Big]de'
\end{equation}
where: 
\begin{equation}
l\equiv\log\lambda L_{\lambda}
\end{equation}
\begin{equation}
m=m(l,\,{\rm FWHM})\equiv\log\frac{M_{\rm BH}}{M_{\odot}}
\end{equation}
\begin{equation}
e=e(l,\,{\rm FWHM})\equiv\log\frac{L_{\rm bol}}{L_{\rm Edd}}.
\end{equation}
Here, the parameters $l_{\rm min}$ and $\sigma_l$, $m_{\rm max}$ and $\sigma_m$, $e_{\rm max}$ and $\sigma_e$ are the minimum luminosity, the maximum mass, the maximum Eddington ratio and the widths of the corresponding distributions. These parameters will be determined in the following via a best fit procedure.

Note that, since the integrals of the $\tilde{P}$ functions diverge, we must restrict their domain to use them as probability densities (e.g., for values of the parameters $l\lesssim l_{\rm min}+3\sigma_l$, $m\gtrsim m_{\rm max}-3\sigma_m$ and $e\gtrsim e_{\rm max}-3\sigma_e$). This doesn't significantly affect the results, because mass, Eddington ratio and luminosity are not independent variables (for example, low mass objects also have low luminosities or high Eddington ratios), hence the derived probability density $P_{\,l,\,{\rm FWHM}}(l,\,{\rm FWHM})$ (Eq. \ref{Ptot}) is essentially insensitive to the shape of $\tilde{P}_m(m)$ at low masses, of $\tilde{P}_e(e)$ at low Eddington ratios or to the shape of $\tilde{P}_l(l)$ at high luminosities.

\begin{figure*}
\begin{minipage}{\textwidth}
\centering
\includegraphics[width=0.6\textwidth, angle=270]{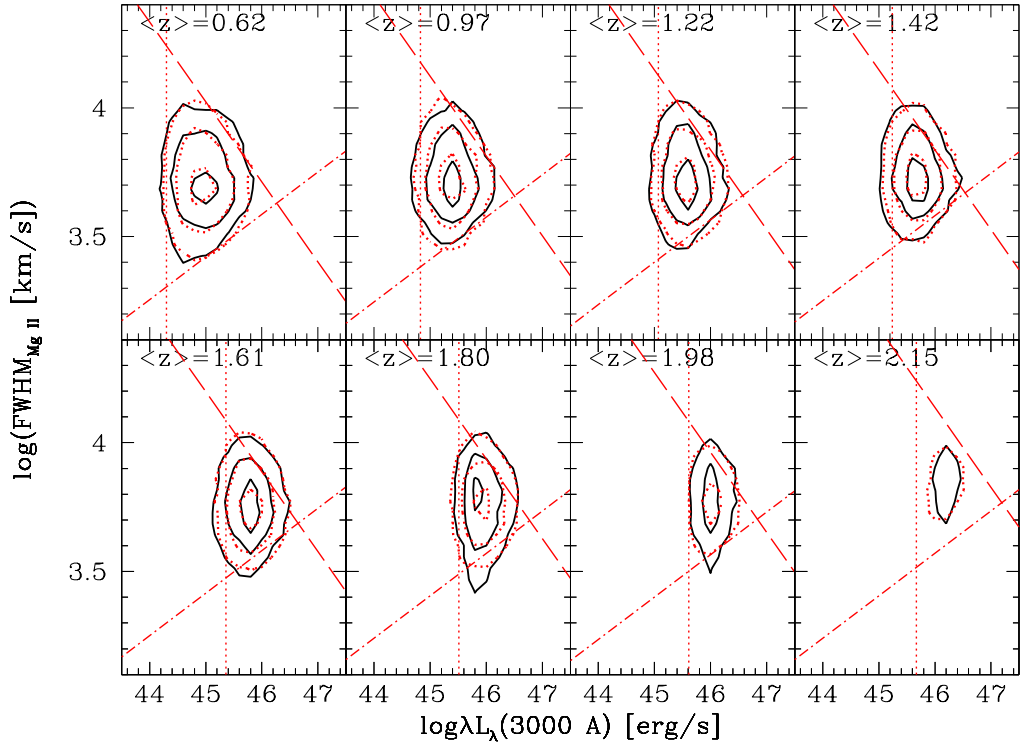}
\caption[]{The 8 panels show the \MgII\ sample in the FWHM--luminosity plane at increasing redshift: solid black contour plot (levels: 20, 90, 250  objects per box, see text) represents the discrete observed distribution of objects. Dotted red contour plot (same levels) shows the discrete distribution of a sample of objects simulated with the Monte Carlo method, adopting the assumed $P_{\,l,\,{\rm FWHM}}(l,\,{\rm FWHM})$ probability density with the best fit parameters. Dotted, dashed and dash-dotted lines represent $l_{\rm min}$, $m_{\rm max}$ and $e_{\rm max}$ respectively.}
\label{graficoz}
\end{minipage}
\end{figure*}
\subsection{Best fit procedure}\label{secfit}
The assumed probability density depends on 6 free parameters, i.e. the minimum luminosity ($l_{\rm min}$), the maximum mass ($m_{\rm max}$) and Eddington ratio ($e_{\rm max}$) and the widths of the corresponding distributions ($\sigma_l$, $\sigma_m$ and $\sigma_e$). We focus on the first redshift bin and determine the free parameters matching with the observed distribution of objects in the FWHM--luminosity plane. In detail, for each choice in the 6-dimension parameter space, the probability density has been constructed, discretized in boxes with $\Delta \log$FWHM=0.04dex 
and $\Delta \log \lambda L_{\lambda}$=0.2dex 
and then normalized to the total number of observed objects, in order to evaluate the expected number of objects in each box ($\Delta\log \lambda L_{\lambda},\,\Delta \log$FWHM). We assumed a Poissonian error (i.e. $\sqrt{n}$) on the observed number of objects in each box. For each choice of the parameters, the expected distribution was compared to the observed distribution in the discrete  $\log \lambda L_{\lambda}-\log$FWHM plane, evaluating the relative $\chi^2$ value. 
The minimum $\chi^2$ determines the best fit parameters.

In order to determine the uncertainties on these values, the same fit procedure was repeated many times comparing the observed distribution  to a set of  simulated distributions of objects, constructed through the Monte Carlo method. This procedure allows an estimate of the error since we sound the underlying probability density only through a finite number of observed objects, the distribution of which in the FWHM--luminosity plane ideally follows Eq. 4 with a certain random dispersion. In detail, given a set of values of the six parameters, we generated $10^7$ points ($\log \lambda {\rm L}_{\lambda},\,\log$FWHM) with uniform probability densities, and then rejected points accordingly to the assumed $P_{L,{\rm FWHM}}(L,{\rm FWHM})$ at given $l_{\rm min}$, $m_{\rm max}$, $e_{\rm max}$, $\sigma_l$, $\sigma_m$ and $\sigma_e$, so that the number of simulated points matches the number of observed objects. We then calculated the root mean square (rms) between the observed and the simulated distributions. This operation was repeated for all the possible combinations of the six parameters (in a reduced phase space around the best fit values). The sextuple which led to the minimum rms gave the so-called Monte Carlo best fit parameters. This procedure was repeated a dozen times, giving as many Monte Carlo best fit values for each parameter, slightly different from one another, but fully consistent with the previous determination. For each parameter, the standard deviation of this set of best fit values was assumed as an estimate of its uncertainty. This uncertainty is much larger than that corresponding to
$\Delta\chi^2=1$.

\begin{table*}
\centering
\begin{minipage}{\textwidth}
\centering
\caption{Best fit values of minimum luminosity, maximum mass, maximum Eddington ratio and widths of the corresponding distributions, with errors and $\chi^2_{\nu}$.  The number of degrees of freedom is $\nu=594$ in the first redshift bin (600 data-points and 6 free parameters) and $\nu=595$ in the others (600 data-points and 5 free parameters). Data that come from a best fit procedure are displayed in boldface.}
\begin{tabular}{@{}ccccccccc@{}}
\hline
\hline
Bin & $<z>$ & $l_{\rm min}$ & $\sigma_l$ &  $m_{\rm max}$ & $\sigma_m$ & $e_{\rm max}$ & $\sigma_e$ & $\chi^2_{\nu}$ \\
\hline
1st &0.62&{\bf 44.30}$\pm0.025$&{\bf 0.26}$\pm0.008$&{\bf 9.18}$\pm0.05$&{\bf 0.31}$\pm 0.003$&{\bf -0.35}$\pm0.02$&{\bf 0.22}$\pm0.003$&3.51\\
\hline
2nd &0.97&44.83                &{\bf 0.23}$\pm0.007$&{\bf 9.35}$\pm0.04$&{\bf 0.31}$\pm 0.006$&{\bf -0.34}$\pm0.02$&{\bf 0.23}$\pm0.003$&4.97\\
3rd &1.22&45.07      	       &{\bf 0.23}$\pm0.005$&{\bf 9.42}$\pm0.05$&{\bf 0.31}$\pm 0.004$&{\bf -0.33}$\pm0.02$&{\bf 0.23}$\pm0.002$&4.63\\
4th &1.42&45.24      	       &{\bf 0.23}$\pm0.008$&{\bf 9.43}$\pm0.05$&{\bf 0.32}$\pm 0.003$&{\bf -0.34}$\pm0.02$&{\bf 0.22}$\pm0.001$&5.11\\
5th &1.61&45.36      	       &{\bf 0.24}$\pm0.006$&{\bf 9.52}$\pm0.04$&{\bf 0.31}$\pm 0.003$&{\bf -0.35}$\pm0.01$&{\bf 0.22}$\pm0.003$&4.36\\
6th &1.80&45.51      	       &{\bf 0.22}$\pm0.005$&{\bf 9.60}$\pm0.05$&{\bf 0.32}$\pm 0.002$&{\bf -0.34}$\pm0.02$&{\bf 0.22}$\pm0.004$&8.44\\
7th &1.98&45.61      	       &{\bf 0.22}$\pm0.005$&{\bf 9.67}$\pm0.05$&{\bf 0.31}$\pm 0.003$&{\bf -0.33}$\pm0.02$&{\bf 0.22}$\pm0.004$&6.84\\
8th &2.15&45.67      	       &{\bf 0.24}$\pm0.006$&{\bf 10.02}$\pm0.05$&{\bf 0.30}$\pm 0.005$&{\bf -0.34}$\pm0.02$&{\bf 0.21}$\pm0.003$&7.21\\
\hline
\label{tab}
\end{tabular}
\end{minipage}
\end{table*}

In the first panel of Fig. \ref{graficoz} we compare the observed distribution with one simulated best fit distribution for the lowest redshift bin. It is apparent that the choice of three simple distributions in luminosity, mass and Eddington ratios describes rather closely the data, giving circumstantial support to the validity of the virial hypothesis on which the theoretical assumptions are based. 

Table \ref{tab} (first line) contains the best fit values of the 6 parameters with relative errors and the reduced $\chi^2$ value (hereafter, $\chi^2_{\nu}$). 
The fact that the $\chi^2_{\nu}$ is larger than 1 is interpreted as due to the choice of an oversimplified distribution. This doesn't influence our results, because our goal is to find a good way to quantify a parameter related to the BH mass (and one related to the Eddington ratio) such that it is not affected by a Malmquist-type bias (i.e. disentangled of the $z$-dependence of the luminosity instrumental limit). In fact, by construction, $m_{\rm max}$ and $e_{\rm max}$ depend neither on the quasar number density nor on the survey flux limit  (see next Section for tests on this statement). 
\subsection{Bias analysis and robustness of the procedure}\label{robust}
\begin{figure}
\includegraphics[width=0.375\textwidth, angle=270]{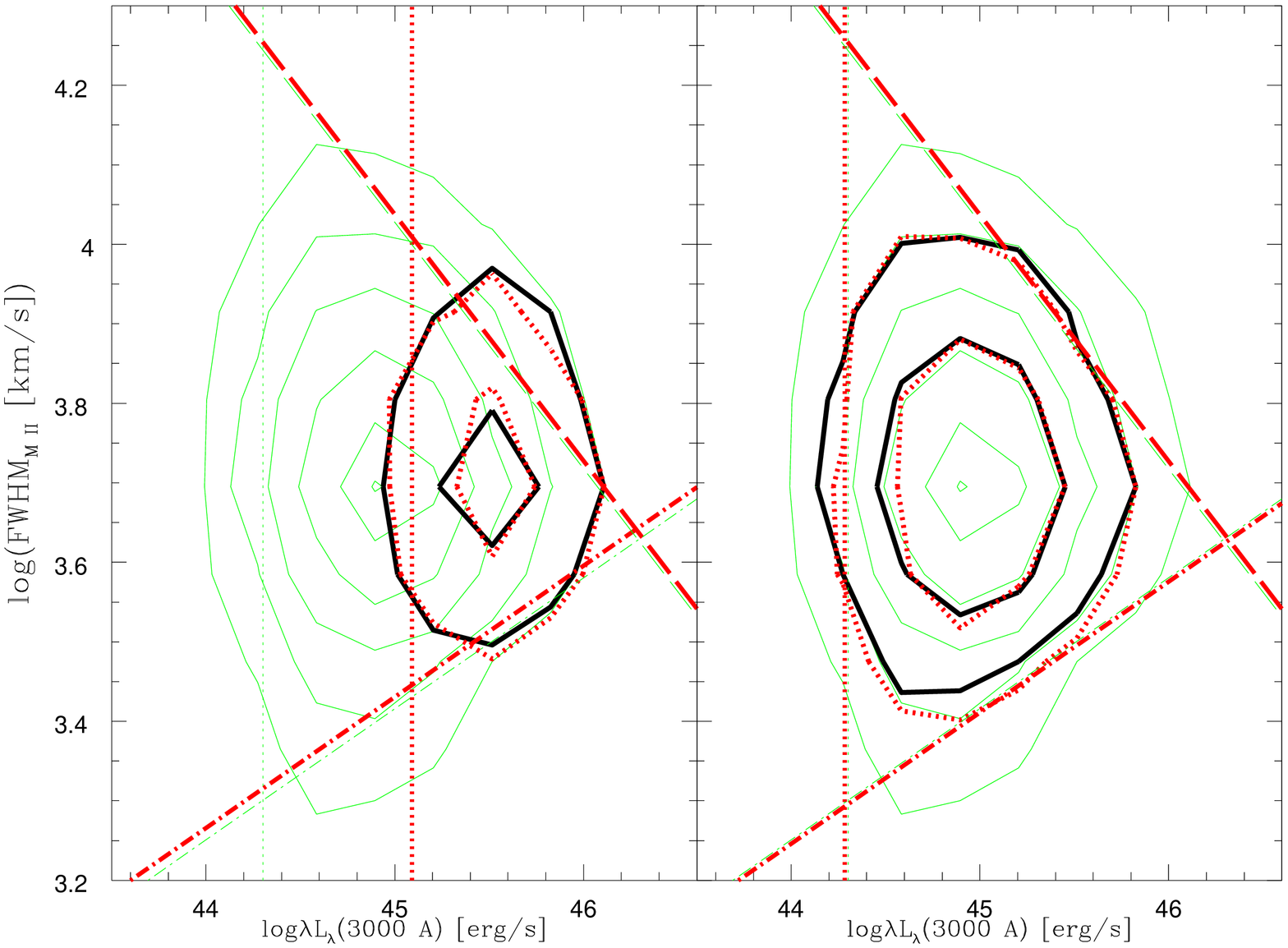}
\caption[]{
The same of Figure~\ref{graficoz} referred to a subsample of the lowest redshift bin objects, selected accordingly to the luminosity cut function of a higher redshift bin (left panel) or randomly selected with $p=0.3$ (right panel). For comparison, green thin lines represent the whole low redshift sample and its best fit $l_{\rm min}$, $m_{\rm max}$ and $e_{\rm max}$ lines.
}
\label{test1}
\end{figure}

The effect of the luminosity cut on the results for $m_{\rm max}$ and $e_{\rm max}$ can be further tested by simulation. 
In order to show that our results are not affected by the instrumental flux limit of the dataset, we selected a subsample from the lowest redshift bin applying the probability function $\tilde{P}_l(l)$ (Eq. \ref{PL}) with a higher luminosity cut, i.e. assuming the $l_{\rm min}$ and $\sigma_{\rm l}$ derived for the third redshift bin in which $<z>=1.22$ (see next Section). This subsample consists of about 1/12 of the objects in the original lowest redshift sample. The fit procedure presented in this paper has been performed again on this subsample. Figure \ref{test1} (left panel) shows that the luminosity cut of a higher redshift bin has negligible effects on the results,
being these values ($m_{\rm max}=9.20$ and $e_{\rm max}=-0.36$) consistent within $1\sigma$ with the best fit parameters derived for the whole sample ($m_{\rm max}=9.18\pm0.05$ and $e_{\rm max}=-0.35\pm0.02$). 

A similar test has been performed to show that $m_{\rm max}$ and $e_{\rm max}$ do not depend on the quasar number density: we re-sampled from the first redshift bin rejecting randomly $2/3$ of the objects, in order to obtain a smaller sample with the same distribution. The fit procedure was then performed on the reduced sample. Again, no significant deviation in the determination of the best fit parameters was observed (see Figure \ref{test1}, right panel). Again, the derived values ($m_{\rm max}=9.20$ and $e_{\rm max}=-0.34$) are consistent within $1\sigma$ with the best fit parameters obtained for the whole sample. These tests show that $m_{\rm max}$ and $e_{\rm max}$ are independent of the quasar number density and of the survey flux limits, and therefore indicate that our procedure is not affected by a Malmquist-type bias. 
\section{Evolution of the QSO population}\label{secevo} 
\subsection{Quasar BH mass and Eddington ratio dependence on redshift}\label{secevoz}
\begin{figure}
\includegraphics[width=0.375\textwidth, angle=270]{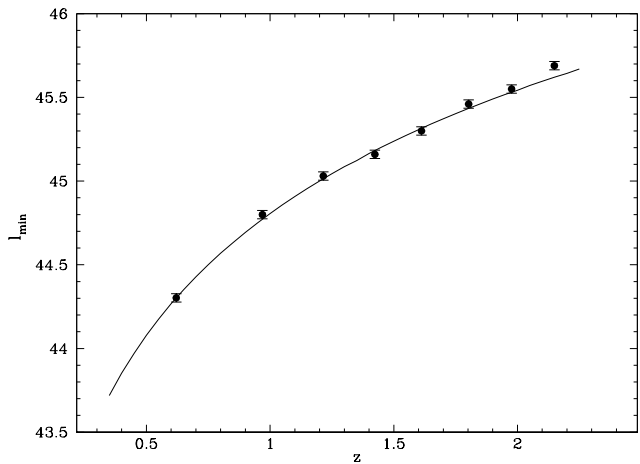}
\caption[]{Black dots represent the best fit parameters $l_{min}$ vs. $z$, 
compared to the values expected from cosmology (solid line).}
\label{plotlumi}
\end{figure}
\begin{figure}
\includegraphics[width=0.375\textwidth, angle=270]{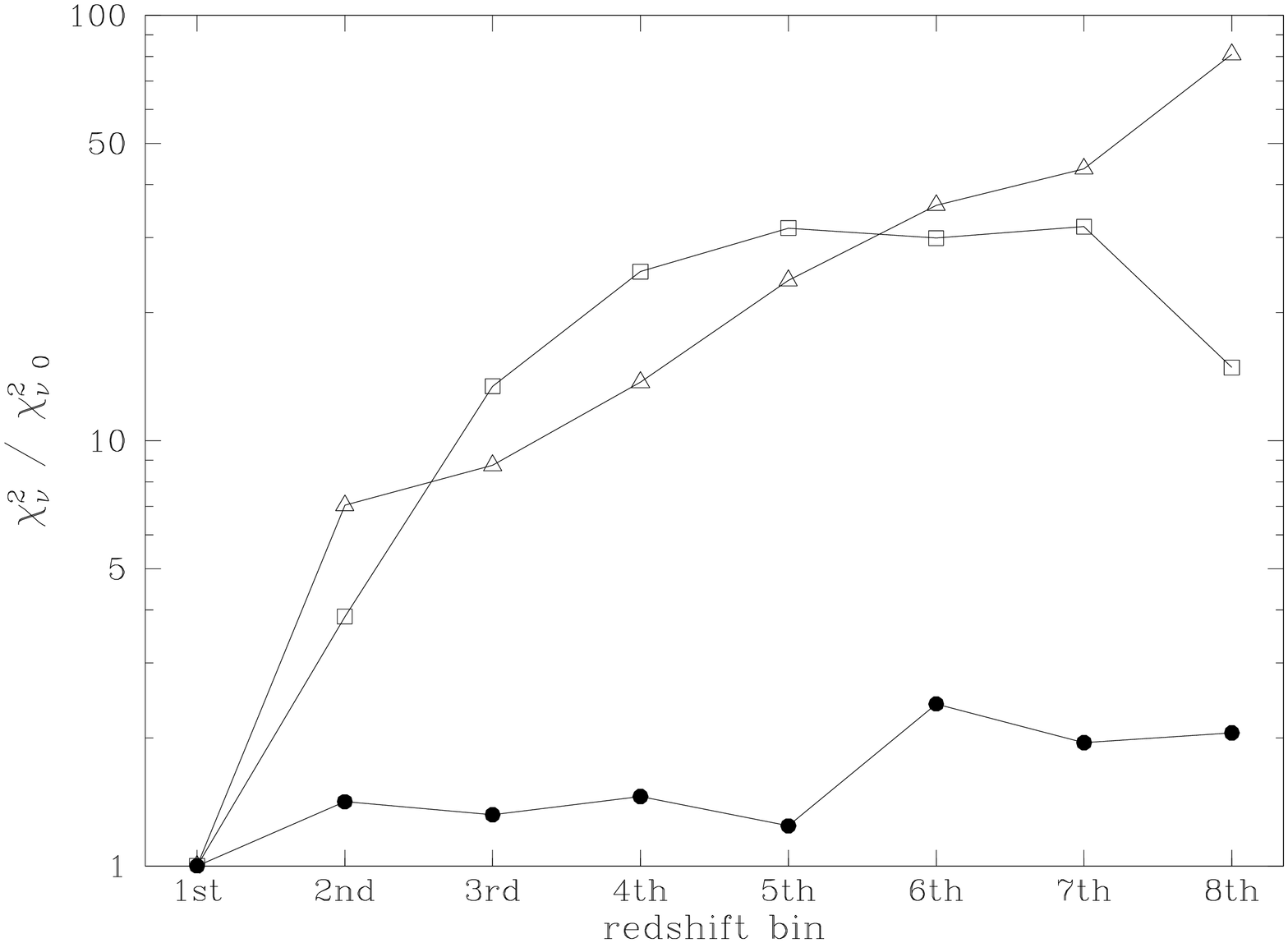}
\caption[]{Normalized $\chi^2_{\nu}$ values of the best fit function. Filled circles refer to the fit procedure described in this work. Open triangles show the $\chi^2_{\nu}$ values that would be obtained assuming that the quasar BH mass is constant with redshift and open squares show the $\chi^2_{\nu}$ values assuming that the BH mass evolves with redshift as proposed by Fine et al. (2006).}
\label{chi2v}
\end{figure}
The fit procedure described above is applied to all the redshift bins, in order to determine the best fit parameters and their uncertainties as a function of redshift. 
In each redshift bin we compared the best fit minimum luminosity with the values inferred through the $z$-dependence of the luminosity distance (see Fig.~\ref{plotlumi}).  
It is apparent that the agreement is very good: apart from the highest redshift bin, where the 
3000 \AA{} continuum is very close to the red edge of the observed spectral range and the 
flux calibration may be unreliable, all the data are consistent with the expectations within $1\sigma$. 
This gives further support to the assumed description of the object distributions in the FWHM--luminosity panels and suggests to repeat the entire procedure assuming that the value of $l_{\rm min}(z)$ is constrained by cosmology.

The same fit procedure is then applied again to all the redshift bins, but  now the dependence on redshift of the minimum luminosity is set by cosmology and $l_{\rm min}$ is no more treated as a free parameter. In each  bin, the $\chi^2_{\nu}$ was evaluated and normalized to the $\chi^2_{\nu { }0}$ value obtained in the first redshift bin, in order to compare the adequacy of the best fit function in the 8 panels. 
Figure \ref{chi2v} shows that these values are almost constant in each redshift bin.
Again, Fig. \ref{graficoz} and Table \ref{tab} show respectively the Monte Carlo simulated distributions compared to the observed distributions of quasars and the best fit values, their errors and relative $\chi^2$ values. 

\begin{figure}
\includegraphics[width=0.5\textwidth, angle=0]{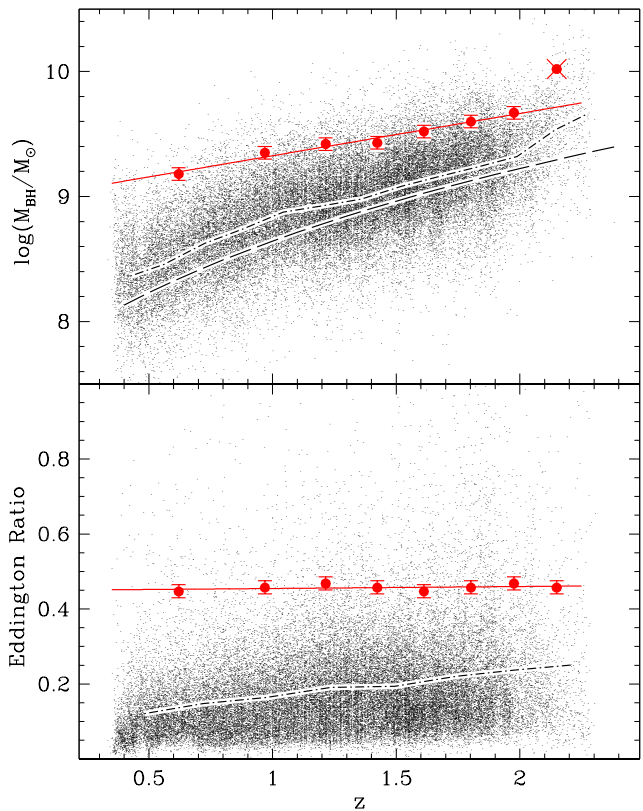}
\caption[]{{\it Upper panel:} small dots are the the virial BH masses of the \MgII\ sample given by Shen et al.~(2008); the dash-dotted line reports the corresponding mean values. Red circles are our estimates of $\log \frac{M_{\rm BH\,(max)}}{{\rm M}_{\odot}}$ and the red solid line is the best fit reported in Eq.~\ref{mmax_z}. The \mbh\ vs.~$z$ dependence proposed by Fine et al.~(2006) is the dashed line.
{\it Lower panel:} small dots are the Eddington ratios for each source of the \MgII\ sample; the dash-dotted line shows the corresponding average. Red circles are the maximum Eddington ratios and the red solid line is the best linear fit reported in Eq.~\ref{emax_z}.}
\label{mez}
\end{figure}

The maximum mass and Eddington ratio values are plotted versus redshift in Figure \ref{mez}.  Note that the proposed \mbh\  $z$-dependence  refers to the active BH population and not to the total supermassive BH mass distribution. Of course, the average mass of the inactive BH population must decrease with increasing redshift. 

A linear fit to the maximum mass values (excluded the highest redshift bin one) gives:
\begin{equation}\label{mmax_z}
\log \frac{M_{\rm BH\,(max)}}{{\rm M}_{\odot}}=m_{\rm max}=0.34(\pm0.02)z+8.99(\pm0.03);
\end{equation}
while the maximum Eddington ratio ($\sim 0.45$) is consistent with no evolution with cosmic time:
\begin{equation}\label{emax_z}
\frac{L_{\rm bol}}{L_{\rm Edd}}\phantom{.}\!_{\rm (max)}=10^{(e_{\rm max})}=0.005(\pm0.006)z+0.45(\pm0.01).
\end{equation}

Assuming that the shapes of \mbh\ and Eddington ratio distributions do not change with redshift, which is suggested by the fact that the value of $\sigma_m$ and $\sigma_e$ are independent of $z$ (see Table \ref{tab}), Eq. \ref{mmax_z} also describes the slope of the $z$-dependence of the mean quasar BH mass, and not only of the maximum mass of the quasar populations. Similarly, the mean (and not only the maximum) Eddington ratio is constant with redshift. 

\subsection{Dependence of the results on the $r_{\rm BLR}-\lambda L_{\lambda}$ calibration}\label{seclll}
\begin{figure}
\includegraphics[width=0.5\textwidth, angle=0]{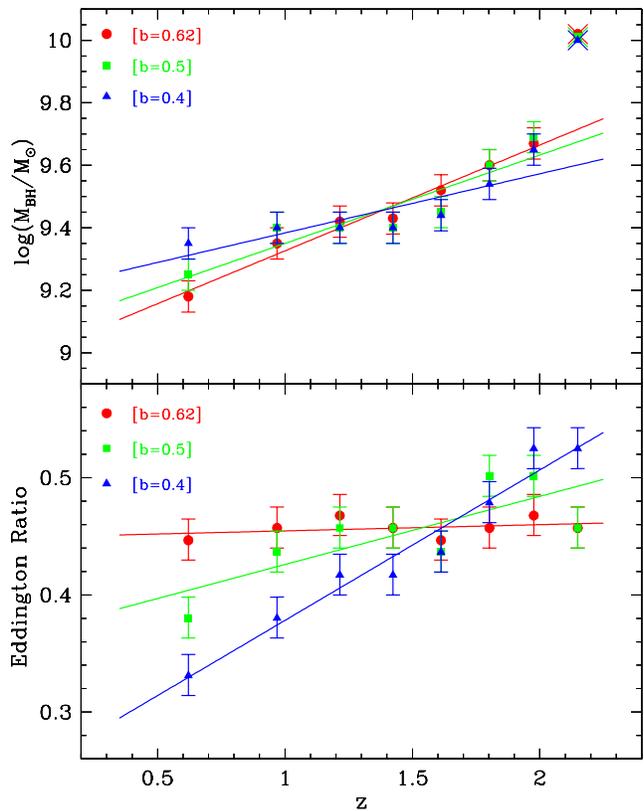}
\caption[]{
BH maximum masses (upper panel) and maximum Eddington ratios (lower panel) as a function of redshift for different values of the luminosity exponent in Eq.~\ref{formulamassa}.
}
\label{test2}
\end{figure}
We tested whether a variation of the luminosity exponent of the virial calibration may affect the relative evolution in $m_{\rm max}$ and $e_{\rm max}$ derived in this paper. The $L-r_{\rm BLR}$ relation assumed in Eq.~\ref{formulamassa} (McLure \& Dunlop 2004) is quite steep, although still consistent with the canonical $r_{\rm BLR}\propto\lambda L_{\lambda}^{\,\,b}$ with $b=0.5$ which is often assumed for idealised photoionisation. In order to quantify the effects that this has on the relative evolution in the maximum mass and Eddington ratio, we reproduced the analysis assuming the exponent on the luminosity term is  $b=0.5$ or $b=0.4$, both of which are consistent to within $2\sigma$ with the McLure \& Dunlop (2004) calibration in which $b=0.62\pm0.14$. 

\begin{table*}
\centering
\begin{minipage}{\textwidth}
\centering
\caption{Best linear fit to the maximum mass and to the maximum Eddington ratio as a function of redshift for various values of the luminosity exponent in Eq.~\ref{formulamassa} ($b=0.62$, $b=0.5$ and $b=0.4$). For each value of $b$, the average value over all the redshift bins of the $\chi^2_{\nu}$ of the best fit probability density (Eq.~\ref{Ptot}) is also given.}
\begin{tabular}{@{}cccccc@{}}
\hline
\hline
    &\multicolumn{2}{c}{$\log \frac{M_{\rm BH\,(max)}}{{\rm M}_{\odot}}=\alpha z+\beta$}&\multicolumn{2}{c}{$\frac{L_{\rm bol}}{L_{\rm Edd}}\phantom{.}\!_{\rm (max)}=\alpha z+\beta$}\\
\\
$b$ & $\alpha$ & $\beta$ & $\alpha$ & $\beta$ & $\langle\chi^2_{\nu}\rangle_z$\\
\hline
0.62&$0.34\pm0.02$&$8.99\pm0.03$&$0.005\pm0.006$&$0.45\pm0.01$&5.6\\
0.5 &$0.28\pm0.05$&$9.07\pm0.08$&$0.06\pm0.02$  &$0.37\pm0.03$&5.8\\
0.4 &$0.19\pm0.05$&$9.19\pm0.07$&$0.13\pm0.01$  &$0.25\pm0.02$&7.5\\
\hline
\label{tabz}
\end{tabular}
\end{minipage}
\end{table*}
Figure \ref{test2} (upper panel) shows that the smaller is the luminosity exponent in the virial calibration, the flatter is the dependence on redshift of the BH masses. The results are only slightly affected by the choice of the $L-r_{\rm BLR}$ relation, since the $z$-evolution determined assuming an exponent of 0.5 is consistent within $1\sigma$ with the previous determination, obtained assuming the McLure \& Dunlop (2004) virial calibration. 
In Table \ref{tabz} we give the best linear fit to the maximum mass as a function of redshift for $b=0.5$, $b=0.4$ and, for comparison, $b=0.62$.

On the other hand, as regards the dependence on redshift of the Eddington ratio, the picture is more delicate. 
This parameter appears to increase significantly with $z$ assuming a flatter $L-r_{\rm BLR}$ relation, while it was found to be constant with redshift within the assumed virial calibration (Eq.~\ref{formulamassa}; see Fig.~\ref{test2}, lower panel). In Table \ref{tabz}, the parameters of the best linear fit to the maximum Eddington ratio as a function of redshift are given for various values of the luminosity exponent ($b=0.5$, $b=0.4$ and, for comparison, $b=0.62$).

Note that the assumption of a flatter  $L-r_{\rm BLR}$ relation leads to an increase of the residuals between the best fit probability density (Eq.~\ref{Ptot}) and the observed quasar distribution. In 
Table~\ref{tabz}, the $\chi^2_{\nu}$ values averaged over $z$ of the best fit probability density are given for $b=0.62$, $b=0.5$ and $b=0.4$. It is apparent that the $\chi^2_{\nu}$ is minimum for $b=0.62$, giving a circumstantial independent support to the index proposed by McLure \& Dunlop (2004, see Eq.~\ref{formulamassa}) and, hence, to Eqs.~\ref{mmax_z} and \ref{emax_z}. 
\subsection{Comparison with previous results}\label{seccomp}
We now compare our results with those obtained by McLure \& Dunlop (2004), Fine \etal\ (2006) and Shen \etal\ (2008), focusing just on the slope of the \mbh\ and Eddington ratio evolution.

Fine et al. (2006),  in order to reduce the effects of the Malmquist bias, concentrated on a subsample of the 2dF quasars with luminosity around $L^*(z)$ at each redshift. They observe a significant dependence of the quasar BH mass on redshift ($\mmbh\propto(1+z)^{3.3\pm1.1}$), but
conclude that their result cannot directly be interpreted as evidence for anti-hierarchical ``downsizing'' because the $z$-dependence they found is strongly dominated by the dependence on redshift of $L^*$. 
For comparison, we repeated the entire fit procedure described above imposing that $M_{\rm BH\,(max)}(z)$ varied as proposed by Fine et al.~(2006).
Figure \ref{chi2v} shows the relative $\chi^2_{\nu}$ values in each redshift bin: the fit appears  inadequate if we assume their results.  Note however that the error given for the evolution of the average BH mass of QSOs by Fine et al. (2004) is quite large, so that their results are consistent with those given here within $1\sigma$.

McLure \& Dunlop (2004) proposed that the observed increase of the quasar BH mass with redshift is entirely as expected due to the effective flux limit of the sample. 
To further test the possibility that the mean BH mass is independent of redshift, we repeated again the fit procedure described above assuming that $M_{\rm BH\,(max)}$ is constant over all the redshift bins. In Figure \ref{chi2v} we plot the relative $\chi^2_{\nu}$, and again the fit is inconsistent with the data, giving further  to evidence for an evolution of quasar populations with $z$. 

McLure \& Dunlop (2004) and Shen et al. (2008), studying the SDSS DR1 and DR5 samples, found that there is a clear upper mass limit of $\sim10^{10}M_{\odot}$ for active BHs at $z>2$, decreasing at lower redshift.
This trend is in good agreement with our results and 
can be explained assuming that the quasar number density peaks at a certain $z_{\rm peak}\sim2-2.5$ and then flattens out (see for example Richards et al. 2006). Around $z_{\rm peak}$, both the high and the low mass end of the quasar BH mass distribution are more populated, so that the observation of very massive objects is likely (while low mass quasars cannot be observed due to the instrumental flux limit). Therefore, the slope of the ``unbiassed'' dependence on redshift of the maximum BH mass is raised below $z_{\rm peak}$ and it is flattened above. This effect  translates in evidence for a limiting BH mass for active BHs at $z>2$, decreasing at lower redshift, that is apparent in all large samples of quasars.

McLure \& Dunlop (2004) observed a substantial increase of Eddington ratios with redshift and a similar trend is apparent from the sample of 2dF $L^*$ quasars of Fine et al. (2006)  after correcting their data for the offset between the \MgII\ and \CIV\ virial mass calibrations (see for example Shen et al.~2008). We suggest that the observed $z$-dependence of Eddington ratios is spurious, and that it is entirely dominated by the dependence on redshift of the average quasar luminosity due to the Malmquist bias. 
\subsection{Discussion of the results}\label{secdisc}
Studying a sample of $\sim50000$ SDSS quasars with $0.35<z<2.25$, we obtained that the maximum mass of the quasar populations increases with $z$, while the maximum Eddington ratio is practically independent of redshift. 

These results are unaffected by the Malmquist bias and may be interpreted as evidence for evolution of the active BH population with redshift. Quasar samples at lower redshift are increasingly dominated by lower mass BHs, i.e.  most massive BHs start  quasar activity before less massive ones. This is indicative of anti-hierarchical ``downsizing'' of active BHs and it is in agreement with recent theoretical predictions by e.g. Merloni, Rudnik \& Di Matteo (2006). 

Our findings may have implications for the study of the co-evolution of supermassive BHs and their host galaxies, even if they cannot be directly interpreted as evidence for evolution of the $M_{\rm BH}-M_{\rm bulge}$ scale relation. There is observational evidence that quasar host galaxies are already fully formed massive ellipticals at $z\sim 2.5$ and then passively fade in luminosity to the present epoch (e.g.~Kotilainen et al. 2007, 2009; Falomo et al. 2008). Within this scenario, our results can be interpreted as an evolution with redshift of the parameter $\Gamma\equiv M_{\rm BH}/M_{\rm bulge}$, which would be 4--5 times larger at $z\sim2$ than today. 

This is in good agreement with the results of Peng et al. (2006), who found that $\Gamma$ is $\sim4$ times larger at $z\sim1.7$ than today in a sample of 11 lensed quasar hosts. 
Our results are also consistent with Salviander et al. (2007), who examined a sample of SDSS quasars finding that galaxies of a given dispersion at $z\sim1$ have BH masses that are larger by $\Delta \log M_{\rm BH}\sim0.2$ than at $z\sim 0$ (see Lauer et al. 2007 for a detailed discussion on the selection bias which may affect these results).
\section{Summary and conclusions}\label{secsum}
Starting from the SDSS DR5 quasar catalogue, we focused on the $\sim50000$ objects for which \MgII\ line widths and 3000\AA\ monochromatic luminosities were available. This sample ($0.35<z<2.25$) was divided in 8 redshift bins. In each bin, the object distribution in the FWHM--luminosity plane was described in terms of a minimum luminosity limit (due to the instrumental flux limit), a maximum mass and a maximum Eddington ratio. 
The assumed probability density was compared to the observed distribution of objects in order to determine the free parameters with a best fit procedure in each redshift bin. Errors on the best fit parameters were determined with Monte Carlo simulations.

We tested the robustness of the procedure through some simulations, and showed that the maximum mass and the maximum Eddington ratio determined in each redshift bin depend neither on the quasar number density nor on the survey flux limit (which is responsible of giving raise to a Malmquist-type bias in the observed dependence on redshift of the mean quasar BH masses). 

We then studied the dependence on redshift of the maximum quasar BH mass and of the maximum Eddington ratio and found clear evidence for evolution of the active BH population with redshift. Over the redshift range studied, we obtained that the maximum mass of the quasar population depends on redshift as $\log (M_{\rm BH\,(max)}/{\rm M}_{\odot})=0.34z+8.99$, while the maximum Eddington ratio is found to be practically independent of redshift. 

This means that QSO samples at lower redshift are increasingly dominated by lower mass BHs, i.e. the more massive a BH is, the earlier it starts quasar activity. Within a scenario in which quasar host galaxies are already fully formed massive ellipticals at $z\sim 2.5$, our results can be also interpreted as an evolution with redshift of the parameter $\Gamma\equiv M_{\rm BH}/M_{\rm bulge}$, which would be 4--5 times larger at $z\sim2$ than today. 
\section*{Acknowledgments}
We wish to thank Yue Shen for providing SDSS quasar spectral measurements before publication. We are grateful to an anonymous referee for constructive criticism, which led to an improvement of the paper.
\begin{footnotesize}

\end{footnotesize}
\label{lastpage}
\end{document}